\documentclass{article}
\usepackage[a4paper, total={6in, 8in}]{geometry}
\usepackage[utf8]{inputenc}
\usepackage{amsmath}
\usepackage{amssymb}
\usepackage{caption}
\usepackage{subcaption}
\usepackage{float}
\usepackage{graphicx}
\usepackage{cancel}
\usepackage{xcolor}
 \usepackage{url}
\usepackage{tikz}
\usepackage[round]{natbib}

\title{Valeriepieris Circles for Spatial Data Analysis}

\author{Rudy Arthur\\
{University of Exeter, Department of Computer Science,}\\
{Stocker Rd, Exeter EX4 4PY}\\
{E-mail:  r.arthur@exeter.ac.uk}
}

\begin{document}

\maketitle

\abstract{
The Valeriepieris circle is the smallest circle that can be draw on the globe containing half of the world's population. The Valeriepieris (VP) circle acts as a spatial median, effectively splitting spatial data into two halves in a unique way. In this paper the idea of the VP circle is generalised and a fast algorithm and corresponding software package to compute it are described. The VP circle is compared to other measures of centre and dispersion for population distributions and is shown to reflect expected differences between countries and changes over time. By studying the VP circle as a function of the included population fraction a new way of representing population distributions is constructed, as well as a mathematical model of its expected behaviour. Finally a measure of population `centralisation' is constructed which measures the tendency of a territory to be dominated by a single population centre or to have a more even distribution of population. 
}

\section{Introduction}

Finding the `centre' of a spatial distribution is common not only in demographics \citep{niedomysl2017using, hall2019population}, but has applications in any field that deals with spatial data, from epidemiology \citep{el2021spatial} to economics \citep{grether2010world, quah2011global}. \cite{rogerson2015new} gives an entertaining account of the often contentious and highly political history of geographic centres in the United States and \cite{murray2018complexities} discusses various `complexities' in defining the centre of any spatial object or distribution. Studying centres over time has been especially common in North America \citep{plane2015tracking, breau2018footsteps, rogerson2021historical} where the location of the population or economic centre of gravity over time shows the westward drift of settler populations in those countries.

There are numerous approaches to finding the `centre' of the world's or a region's population. One approach, taken by the US Census bureau \citep{uscentres}, is to compute the \textbf{centre of population} from
\begin{align}\label{eqn:cpop}
\hat{\phi} &= \frac{\sum_i w_i \phi_i}{w_i} \\ \nonumber
\hat{\lambda} &= \frac{\sum_i w_i \lambda_i \cos(\phi_i) }{w_i \cos(\phi_i) }
\end{align}
Where the sum is over $i$ areas with population $w_i$ and latitude and longitude $(\phi_i, \lambda_i)$ to yield $(\hat{\phi}, \hat{\lambda})$, the latitude and longitude of the centre. \cite{barmore1992we} notes that this gives the point at which a flat map would balance if the population was placed on a sinusoidally (Sanson–Flamsteed) projected map with central meridian at longitude $\hat{\lambda}$. \cite{barmore1992we} suggests the centre of population be defined instead as the balance point on a map projected using the Azimuthal Equidistant projection centred at the centre of population. This would keep distances of populations to the centre undistorted but must be computed iteratively.

The dependence of the above on the map projection has provoked alternative definitions. An influential one is given by \cite{aboufadel2006new} who locate population mass using three dimensional vectors originating at the Earth's centre and compute the weighted average vector
\begin{align}\label{eqn:cpop3}
    \hat{x} &= \frac{\sum_i w_i \vec{x}_i}{ \sum_i w_i} \\ \nonumber
    \bar{x} &= \frac{\hat{x}}{|\hat{x}|} 
\end{align}
Here $\vec{x}_i$ are the vectors locating each of the populated areas and $\hat{x}$ is the population weighted average vector. \cite{barmore1992we} and others e.g. \cite{grether2010world} note that the weighted average $\hat{x}$ produces points not on the surface of the earth, though $\bar{x}$ will be on the surface. I will refer to this point as the \textbf{3d centre of population}. In practice \citep{rogerson2021historical} finds that this point and \cite{barmore1992we}'s iterative Azimuthal centre are very similar, though the 3d point is much easier to compute.

There are other definitions of centre which do not use a weighted sum of population. The intersection of the parallels of latitude and longitude that divide the target population in half is called the median centre or cross-median \citep{plane2015tracking}. This definition depends on the orientation of the latitude and longitude grid and the longitude median is undefined for the entire globe. The \textbf{geometric median} is another possible definition - this is the point for which the sum of distances to each individual is minimised, however it is somewhat complicated to compute \citep{beck2015weiszfeld}. The geographic centre is a related idea - this the point where the sum of squared distances from the center to all points in the region is minimised. \cite{rogerson2015new} provides a method to calculate this, which is similar to the Azimuthal iterative method of \cite{barmore1992we}.

As pointed out by \cite{barmore1992we}, we want a statistic that indicates where people are, not where they could easily travel to. Likewise we want a statistic that doesn't depend on arbitrary map constructions - including how we draw grid lines or how we project onto the plane. The centre of population fits most of these criteria, however the centre of population it is often far from where most people actually are. The centre of the US population is in southern Missouri, the global centre of population is in the Arabian desert (see the following sections for the exact coordinates). For showing trends over time the actual location of the centre may not be so important, only the relative change from year to year matters, however as a `representative point' the centres above all leave something to be desired.

\subsection{The Valeriepieris Circle}

The Valeriepieris circle\footnote{Valeriepieris is the Reddit username of Ken Myers who first suggested the idea there: \url{https://www.reddit.com/r/MapPorn/comments/1dqh7d/after_seeing_a_recent_post_about_the_population/} } is the circle of smallest radius containing half the Earth's population. The idea gained popular attention e.g. \citep{washingtonval} and some academics, notably economist Danny Quah \citep{quahtight, quah2016ordering}, discussed the idea. The Valeriepieris circle for the whole earth is centred in the north of Yunnan province, China with a radius of around 3300km. The radius is remarkably small due to the very high populations of northern India, southern China and the rest of south east Asia which together constitute half the world population.

The idea behind the Valeriepieris (hereafter VP) circle can naturally be generalised. The VP circle is the answer to the question: \emph{What is the smallest circle containing at least a fraction $f$ of the population of the area $c$?} The original VP circle has $f=0.5$ and $c$ equal to the whole Earth, however we can use any fraction $f$ and any populated area: Europe, England, Exeter etc. for $c$. 
I will refer to the radius of this circle as the \textbf{VP f-radius} for $c$, denoted by $R(f)$. 
Similarly call the centre of the circle the \textbf{VP f-centre}, denoted $( \phi_{VP}(f), \lambda_{VP}(f) )$. I will refer to the special case with $f=0.5$ as \emph{the} VP circle and if the $f$ value is not specified it is assumed to be $0.5$ by default.

The VP radius can be used as a measure of population dispersion. A small radius indicates a population concentrated around a central point, a large one indicates a more uniform spread. Some other measures of population dispersion are discussed in \cite{rogerson2021historical}. The simplest and only one I consider is \textbf{Bachi standard distance} \citep{bachi1963standard}
\begin{equation}\label{eqn:bachi}
   s(y) = \sqrt{ \frac{\sum_i w_i d_i(y)^2 }{w_i} },
\end{equation}
where $d_i$ is the distance between the location $i$ and the centre $y$. 

In this paper I will argue that the VP centres and radii provide an interesting and useful summary of spatial data while avoiding some of the problems mentioned above. In Section \ref{sec:finding} I give a method to compute VP circles for any area and compare the VP circle to some of the other centre definitions mentioned above. In Section \ref{sec:cpop} I look at the VP radius as a function of time, a popular use case for population centre statistics. Section \ref{sec:profiles} looks at VP circles as a function of $f$. This yields an interesting statistic that reduces spatial population distributions from two to one dimension which can be fit with a simple amthemathical model. I also give an statistic which characterises patterns of population density in a novel way. I conclude in Section \ref{sec:conc}.

\section{Finding Valeriepieris Circles}\label{sec:finding}

I will use gridded population data from SEDAC \citep{sedac}. This data is available at grid resolutions from 1 degree ($\sim 111 \text{km}$) to 30 seconds ($\sim 1 \text{km}$) for the years 2000, 2005, 2010, 2015 and 2020. Having data on a regular grid is not necessary for finding VP circles, but makes a number of computational efficiencies possible. The algorithm below will find a VP circle, with any kind of population data, by an exhaustive search.

\begin{enumerate}
\item Compute the total population $P$ of $c$ and the target population $t = fP$. Choose a starting point $(\phi_0, \lambda_0)$.
\item Using a binary search, grow a circle centred at $(\phi_0, \lambda_0)$ to find the minimum radius where the population within the circle is $\ge t$. Call this radius $R_0$.
\item Shift the circle to the next location and calculate the population within a distance $R_0$ from this point. 
\item If the population is $\ge t$ find the radius $R_1$ (again by binary search) such that $R_1$ is the smallest possible radius where the population is still $\ge t$.
\item After checking all the locations, the final radius $R(f)$ will be the VP f-radius and the grid points where the population within the circle is $\ge t$ will be the VP f-centres.
\end{enumerate}
I refer to VP centres, plural, because in some cases, typically when $f$ is very small and the grid is coarse, there may be multiple minima. In practice, for any reasonably fine grid and $f \gtrapprox 0.05$, this doesn't happen and the VP circle found by the above process is unique.

The binary search proceeds in the obvious way: starting from , say, $(r_{min}, r_{max}) = (0, R_c)$, where $R_c$ is the greatest distance between any two points in $c$, compute the population, $P_0$, within a distance $r_0 = \frac{r_{min} + r_{max}}{2}$ from the centre. If $P_0 < t$ then set $r_{min} = r_0$ otherwise set $r_{max} = r_0$ and recalculate the population. The division is repeated until $r_{max} - r_{min} < \epsilon$, i.e. we have narrowed down the radius to within some threshold $\epsilon$. Once this happens set $R = r_{max}$. I have chosen the threshold $\epsilon$ to be $1 \text{km}$ which is smaller than the grid resolutions I will use and means this process usually converges in fewer than 10 iterations.

When the data forms a regular grid the search can be made significantly more efficient:
\begin{itemize}
\item When shifting the centre to a neighbouring grid point, only a small number of population counts change. By keeping track of which grid points are on the edges of the circle we can add and remove only the necessary data.
\item Using the symmetry of the sphere, distances only need to be recalculated when the latitude of the centre point changes, otherwise, at fixed latitude, the same distance matrix can be shifted and reused.
\item If the VP circle is first found on a coarse grid we can narrow down the search area significantly e.g. only search within, say, $\pm 5$ degrees of the coarse centre on the fine grid.
\item Since the distance calculations are the same every time, if multiple $f$ values are required, do all of them at the same time.
\end{itemize}
I have implemented the above algorithm in Cython \citep{behnel2011cython} and made it available as a Python package\footnote{ \url{https://pypi.org/project/valeriepieris/} }. On a single core (Intel i7 1.9 GHz) using the SEDAC 15 minute resolution data, which covers the globe with a $720 \times 1440$ grid, the VP circle takes about 16.5 seconds to compute. If the coarse grid is used first to roughly locate the centre, the whole process is significantly quicker, completing in less than 1 second. I have also implemented a number of other popular methods to calculate population centres
\begin{enumerate}
\item \textbf{Centre of population}, see equation \ref{eqn:cpop}
\item \textbf{3d Centre of population}, see equation \ref{eqn:cpop3}
\item \textbf{Geometric median}
\end{enumerate}
I calculate geometric median using Weiszfeld’s Method \citep{weiszfeld} modified after \cite{vardi2001modified} to work when the centre is at a grid point.

\begin{table}
\centering
\begin{tabular}{|c|c|c|c|}
\hline
Method & (Lat, Lon) & Bachi standard distance (km) \\ \hline
Centre of population & $(22.125 , 51.375)$ & 6583 \\
3d Centre of population& $(36.625 , 66.875)$ & 6364 \\
Geometric median & $(24.625 , 72.125)$ & 6629 \\ \hline \hline
 & $(\phi_{VP} , \lambda_{VP})$ & $R$ (km) \\ \hline
VP 0.5-circle & $(28.375 , 100.625)$ & 3386 \\ \hline
\end{tabular}
\caption{Positions of the various centres and radii for the globe. All lat, lon values have been `snapped' the the nearest grid point and radii are only reported to the nearest km.}
\label{tab:vpglobe}
\end{table}
\begin{figure}
    \centering
    \includegraphics[width=\textwidth]{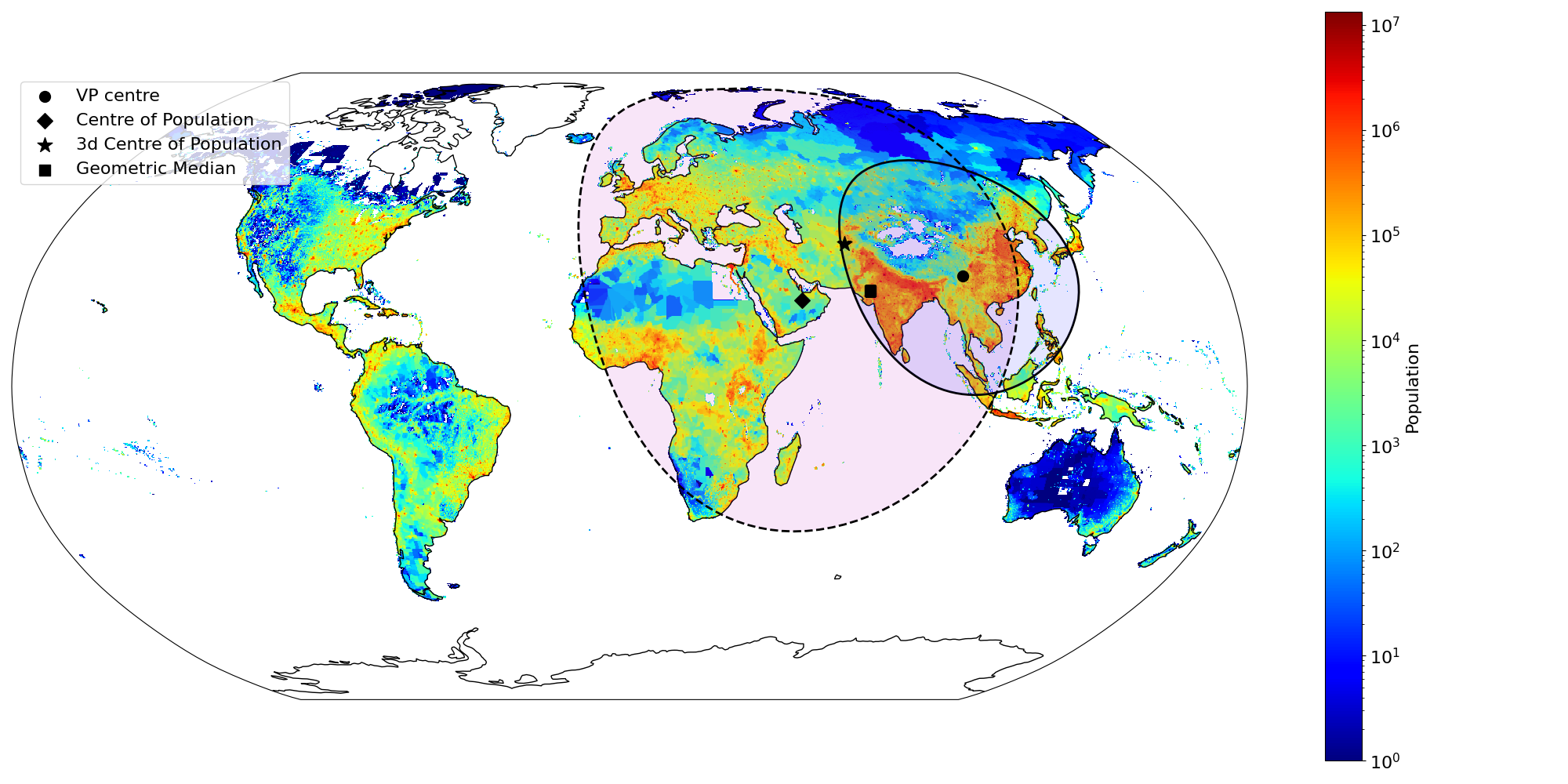}
    \caption{The Valeriepieris circle (small circle, solid outline, blue fill) for the entire globe. The centre is positioned roughly half-way between India and China, see Table \ref{tab:vpglobe} for the exact location. Other population centres are shown, as well as a circle of radius equal to the Bachi standard distance with origin at the Centre of population (large circle, dashed outline, pink fill). Using SEDAC 15 minute resolution, $\sim 30 \text{km}$, data.}
    \label{fig:vpglobal}
\end{figure}

As found in the original Reddit post, the VP circle for the entire globe is centred between India and China, see Figure \ref{fig:vpglobal}. The exact position and radius are given in Table \ref{tab:vpglobe} which also gives co-ordinates of the other centres. As most of the world's population is in south east Asia, all 4 centres cluster around there, and interestingly, all 4 centres are within 10 degrees of latitude of each other. The Centre of population, and 3d Centre of population are furthest west, influenced by sizeable populations in Africa, Europe and the Americas. The geometric median is closest to the VP centre, which is expected - the VP circle is also a kind of spatial median, half the data is inside the circle and half outside and median statistics are less influenced by the extremes of the distribution they summarise. The VP radius is about half of the Bachi standard distance.

\begin{figure}
    \centering
    \includegraphics[width=\textwidth]{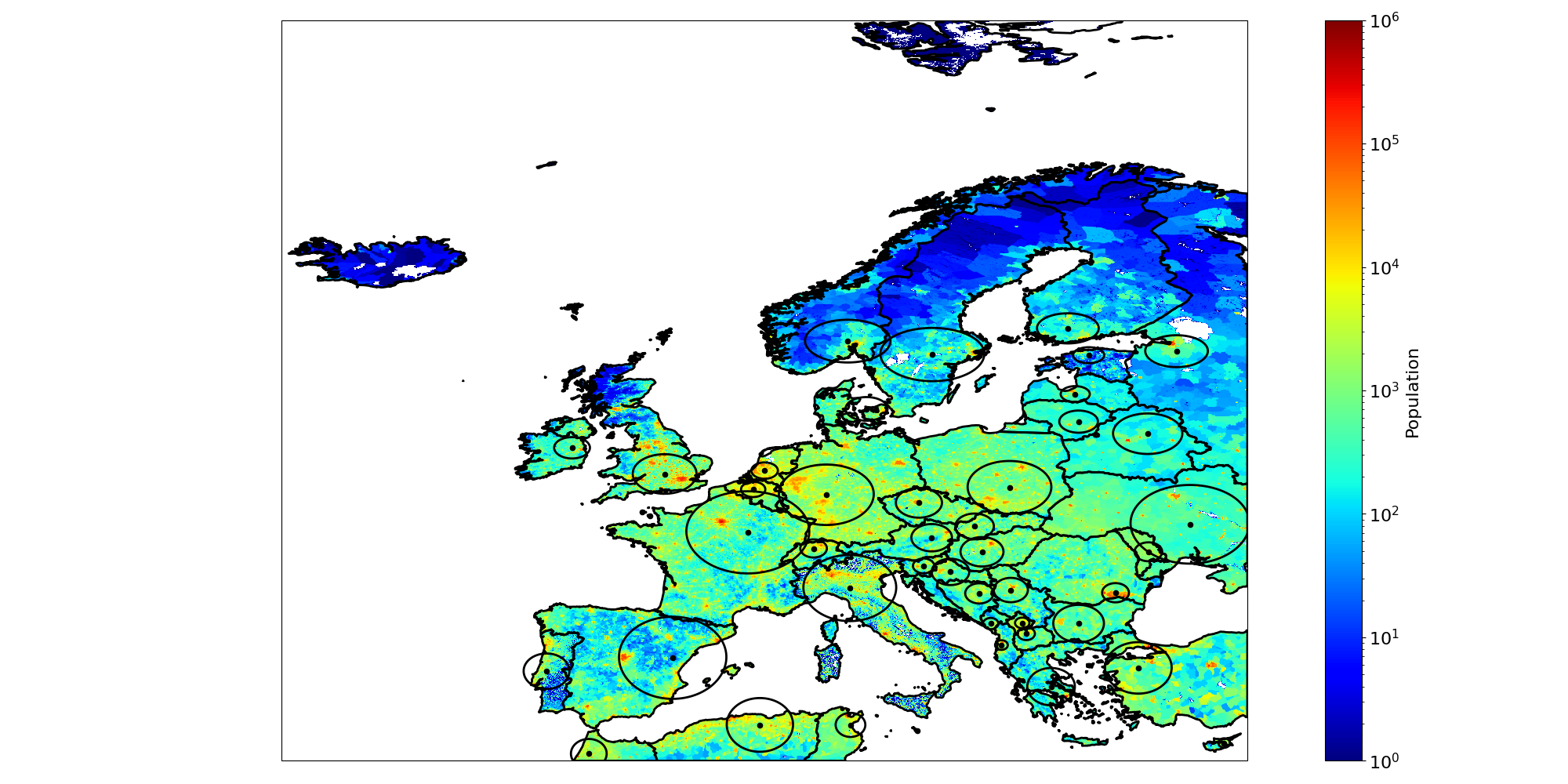}
    \caption{The Valeriepieris circles for countries in `Europe' here eqivalent to a bounding box with lower left lat, lon $(34,-25)$ and upper right lat, lon $(80,35)$. Only the portions of countries that intersect with this bounding box are used in the VP calculations (e.g. only western Russia). Each country is treated individually, so even if the VP circle crosses a border, population in the other country is not counted. Using SEDAC 2.5 minute resolution, $\sim 5 \text{km}$, data.}
    \label{fig:vpeurope}
\end{figure}

The VP circles for a number of countries in Europe are shown in Figure \ref{fig:vpeurope}. Each one takes a few seconds to compute, in fact the bottleneck is usually the intersection of the polygon data (obtained from \citep{WBpoly}) with the population grid. Most country's VP centres are close to that country's capital/largest city. Most are quite a bit smaller than their country, some very much so e.g. Iceland's VP radius is only 15 km. We will look at some specific countries in more detail in the next section and study the behaviour of the VP radius as a function of $f$ in Section \ref{sec:profiles}.

\section{Centres of Population over Time}\label{sec:cpop}


One of the most prominent uses of population centres has been to track the change in population in the US over time, in particular its westward movement \citep{rogerson2015new, plane2015tracking, rogerson2021historical}. Population centres of the UK and its constituent countries have also attracted interest. \cite{dorling1995population} suggest that the population centre of Great Britain has been moving south at about 100 meters per year.

\begin{table}
\centering
\begin{tabular}{|c|c|c|}
\hline
\multicolumn{3}{|c|}{Continental US}   \\ \hline
\hline
Year & $( \phi_{VP}, \lambda_{VP} )$ & $R$ (km) \\ \hline
2000 & $(39.7708, -80.6875)$ & 863 \\
2005 & $(39.4792, -80.6042)$ & 878 \\ 
2010 & $(39.7292, -80.9375)$ & 901 \\ 
2015 & $(39.4792, -81.2708)$ & 923 \\ 
2020 & $(38.3958, -80.6875)$ & 949 \\ \hline
\end{tabular}
\begin{tabular}{|c|c|c|}
\hline
\multicolumn{3}{|c|}{Great Britain}   \\ \hline
\hline
Year & $( \phi_{VP}, \lambda_{VP} )$ & $R$ (km) \\ \hline
2000 & $(51.8124, -1.1458)$ & 138 \\
2005 & $(51.8124, -1.1458)$ & 136 \\ 
2010 & $(51.8124, -1.1458)$ & 135 \\ 
2015 & $(51.8124, -1.1458)$ & 134 \\ 
2020 & $(51.8124, -1.1458)$ & 132 \\ \hline
\end{tabular}
\begin{tabular}{|c|c|c|}
\hline
\multicolumn{3}{|c|}{Mongolia}   \\ \hline
\hline
Year & $( \phi_{VP}, \lambda_{VP} )$ & $R$ (km) \\ \hline
2000 & $(47.7292, 104.4792)$ & 227 \\
2005 & $(48.7708, 105.7708)$ & 167 \\
2010 & $(48.4375, 105.6875)$ & 134 \\
2015 & $(48.0625, 107.4792)$ & 83 \\
2020 & $(47.9792, 106.8542)$ & 34 \\ \hline
\end{tabular}
\caption{VP centre and VP radius for 3 areas over time. All lat, lon values have been `snapped' the the nearest grid point and radii are only reported to the nearest km.}
\label{tab:vptime}
\end{table}

\begin{figure}
    \centering
    \includegraphics[width=\textwidth]{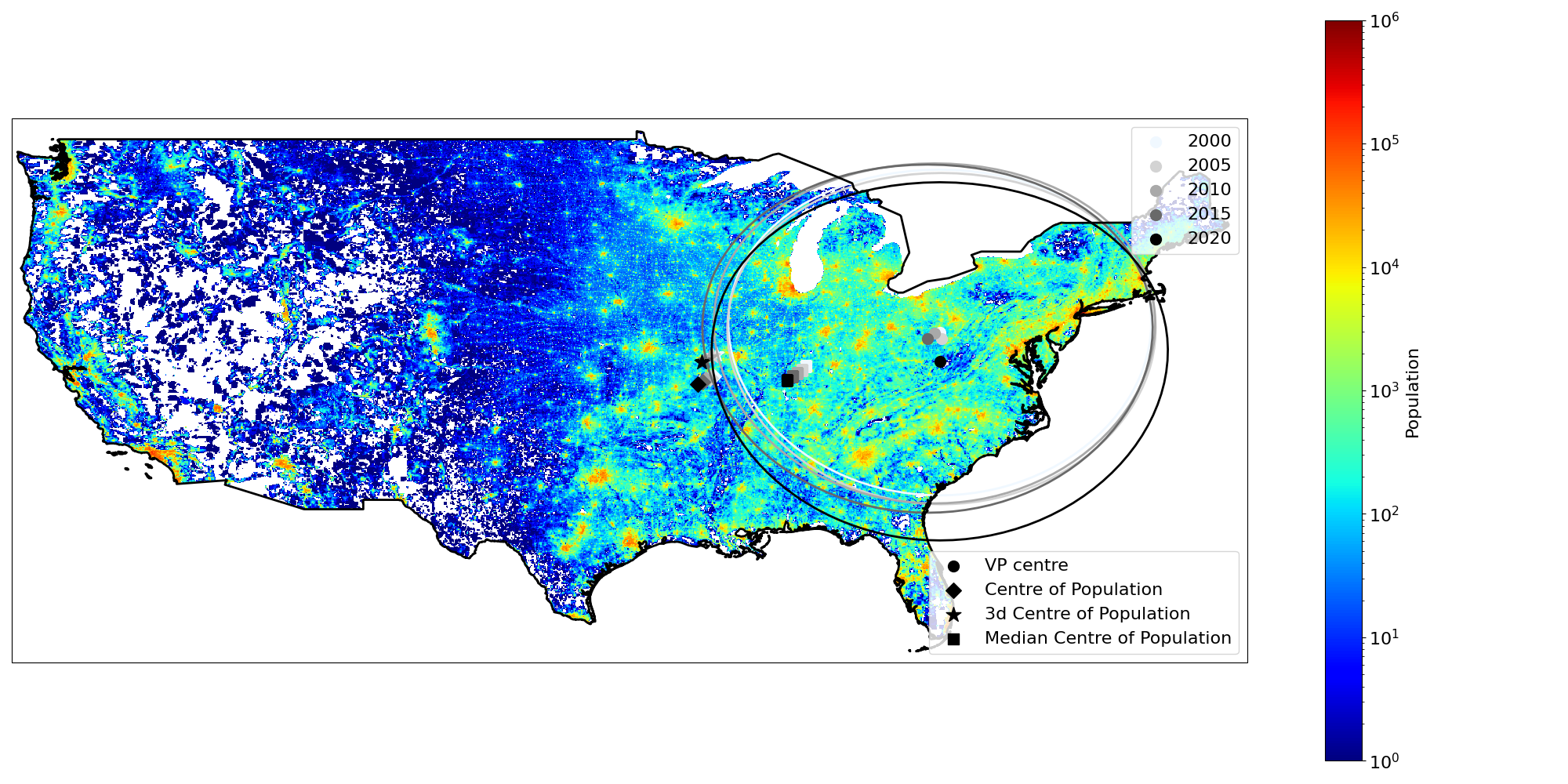}
    \caption{Valeriepieris circles for the continental US between 2000 and 2020 overlayed on 2020 population data. Also shown are the other centres of population in the same period. Using SEDAC 2.5 minute resolution, $\sim 5 \text{km}$, data.}
    \label{fig:vpus}
\end{figure}
Figure \ref{fig:vpus} shows the VP circles, Centre of Population, 3d Centre of Population and Geometric median for the continental United States (i.e. not including Alaska and Hawai'i). First note again that the VP centre is significantly further east and closer to the main `mass' of population than the others (it is near Craigsville West Virginia). The VP radius is around 950km (see Table \ref{tab:vptime} for precise values) and encompasses the major cities of the north east, great lakes and eastern sun belt. All the centres show a gradual southwest drift, though between 2015 and 2020 the VP centre makes a sudden southward jump while moving back east slightly. This is less significant in light of the growing VP radius, which increased by around 90km over two decades. The VP circle puts the centre of the US in the east and with New York at the north east boundary, Chicago at the north west and Atlanta at the south. The median American lives in this circle.  The changing demographics of the US are reflected not as much by the position of the centre as by the growing VP radius. Proportionally higher population growth in the south and west means that progressively larger radii are required to `capture' 50\% of the population, even if the centre doesn't move as far west.

\begin{figure}
    \centering
    \includegraphics[width=\textwidth]{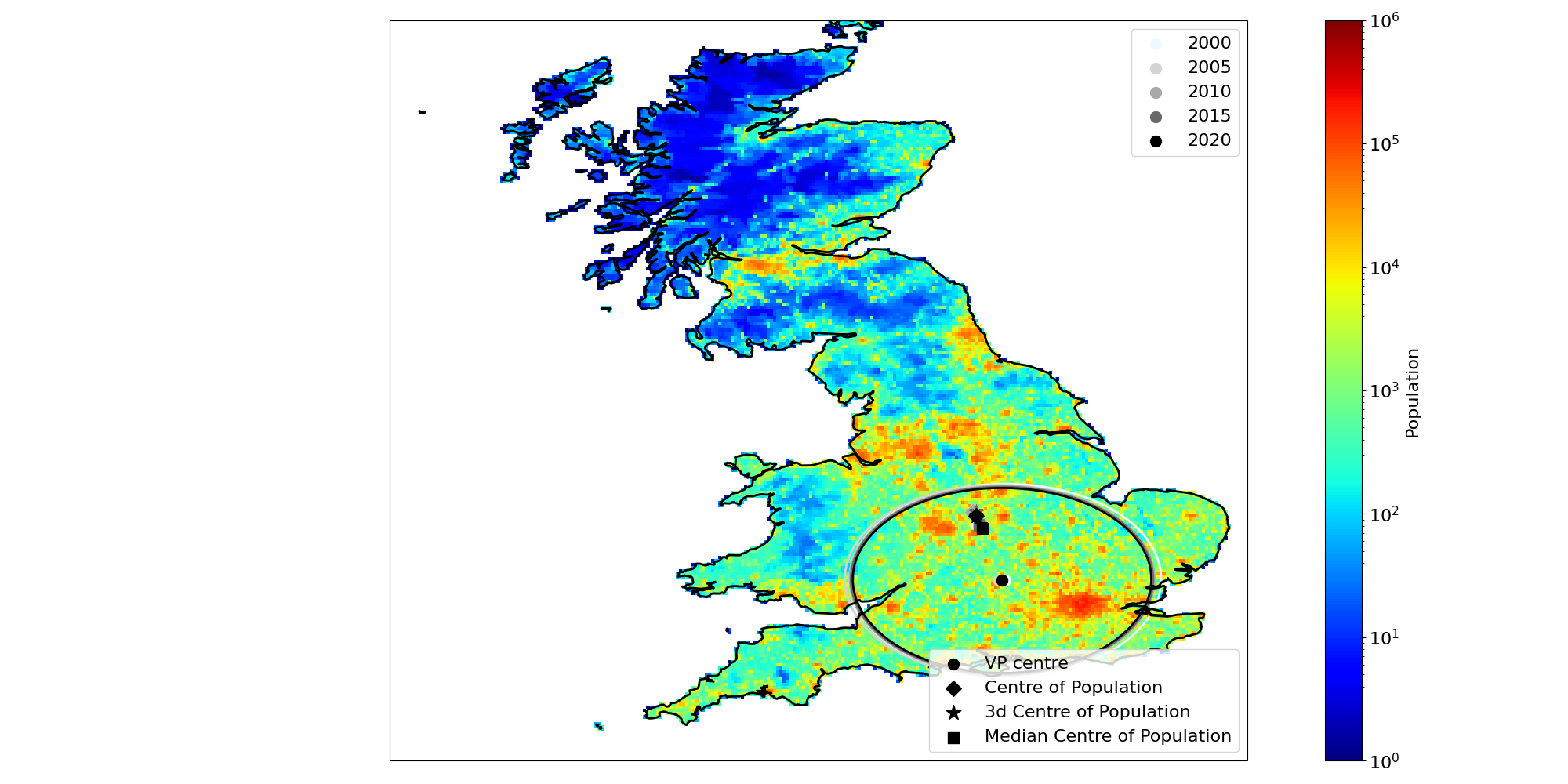}
    \caption{Valeriepieris circles for Great Britain between 2000 and 2020 overlayed on 2020 population data. Also shown are the other centres of population in the same period. Using SEDAC 2.5 minute resolution, $\sim 5 \text{km}$, data. }
    \label{fig:vpuk}
\end{figure}
The US is an unusual case, with heavily populated east and west coasts separated by a large expanse of sparsely populated territory. Figure \ref{fig:vpuk} shows the UK over the same period. Compared to the US there is very little change over time. None of the centres move very much, in fact, the VP centre is located in the same grid cell for all dates. The radius decreases slightly, by about 2 km every 5 years, see Table \ref{tab:vptime} for precise values. This is generally in line with the rather slow evolution of the UK population in the 20th century as reported by \cite{dorling1995population}.

\begin{figure}
    \centering
    \includegraphics[width=\textwidth]{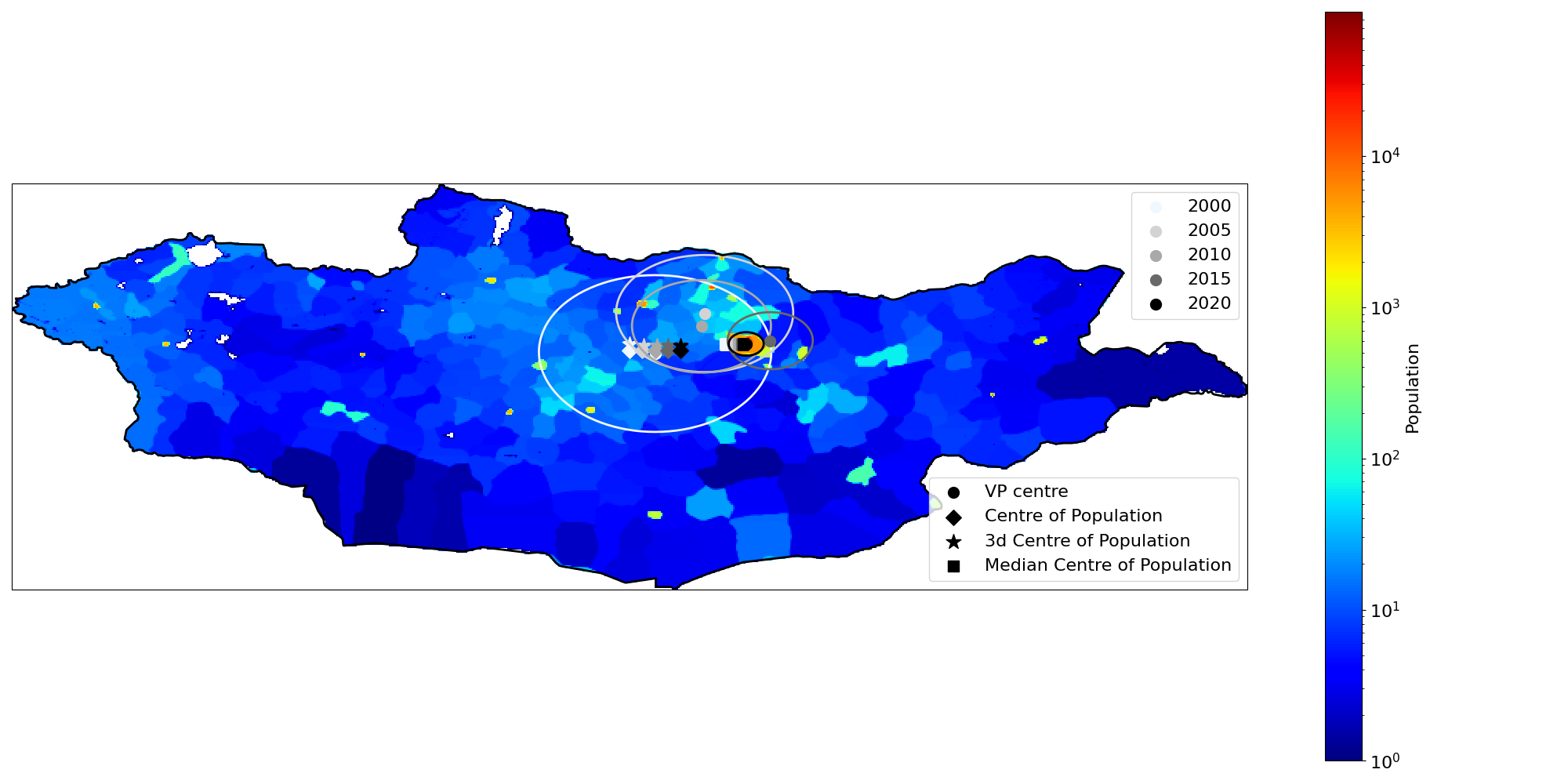}
    \caption{Valeriepieris circles for Mongolia between 2000 and 2020 overlayed on 2020 population data. Also shown are the other centres of population in the same period. Using SEDAC 2.5 minute resolution, $\sim 5 \text{km}$, data. }
    \label{fig:vpmongolia}
\end{figure}
The US is the country with the largest increase in VP radius from 2000 to 2020 at 86km, indicating a more even dispersal of population across the territory. At the opposite extreme, the largest reduction in VP radius, indicating population concentration, is found in Mongolia whose VP radius has reduced by 193km, see Figure \ref{fig:vpmongolia} and Table \ref{tab:vptime} for precise values. Mongolia has experienced huge rural to urban migration in recent years \citep{mongolia} which is shown quite cleanly by reduction in VP radius. The examples of the US and Mongolia suggest that changes in VP radius over time are indicative of changing demographic patterns.

\section{Valeriepieris Profiles} \label{sec:profiles}

In this section I will look at how the VP f-radius changes as a function of $f$, the fraction of the population included in the circle. To build some intuition it will be useful to consider a toy example. Let $R(f)$ be the VP f-radius for some area. We must have $R(0) = 0$. $R(1)$ corresponds to the radius of the smallest circle that encompasses the entire population and so is determined only by the shape of the populated area. Imagine a circular island of radius $R_I$ with a radially symmetric population density $\rho(r)$. The population at $R$ is
\begin{align*}
P(R) = 2\pi \int_0^{R} r dr \rho(r)
\end{align*}
Set \[
\rho(r) = 
\begin{cases}
    \frac{\rho_0 R_0^{2-a}}{ \left( r+R_0 \right)^{2-a} } ,& \text{if } r\leq R_I\\
    0,              & \text{otherwise}
\end{cases}
\] 
The population density equals $\rho_0$ at the centre, $r=0$, and increases or decreases like a power of the radius as we move away from the center up to some max or min at the boundary of the island, $R_I$. We then have
\begin{align*}
P(R) = \rho_0 R_0^{2-a} \pi \int_{0}^{R} r \left( r+R_0 \right)^{a-2}  dr =\rho_0 R_0^{2-a} \pi (r+R_0)^a \left( \frac{1}{a} - \frac{b}{(r+R_0)(a-1)} \right) \bigg\vert_0^R
\end{align*}
The special cases of $a=0$ and $a=1$ have to be treated separately, where the integral gives a log. Ignoring this complication, if the total population is $P = P(R_I)$ then solving the above for $R$ with $P(R) = fP$ gives the VP f-radius. For large R we have approximately
\begin{align*}
(R(f)+R_0)^a \simeq \frac{afP}{\rho_0 R_0^{2-a} \pi}
\end{align*}
or even more approximately: $R \sim f^{1/a}$. It is convenient to normalise by $R(1)$ to create a function that is bounded between 0 and 1, which allows a direct comparison of places with very different areas. Define
\begin{align}
\tau(f) = \frac{R(f)}{R(1)} 
\end{align}
 The special case of a uniform population density corresponds to $a=2$ and can be solved exactly where the (normalised) VP f-radius grows like the square root of $f$. For $a < 2$ the radius grows more slowly, since the population is decaying more rapidly than the circle's area grows as a function of $R$. I will call the normalised VP f-radius, considered as a function of the population fraction, $\tau(f)$, the \textbf{VP profile} for some area c.

\begin{figure}
    \centering
    \includegraphics[width=\textwidth]{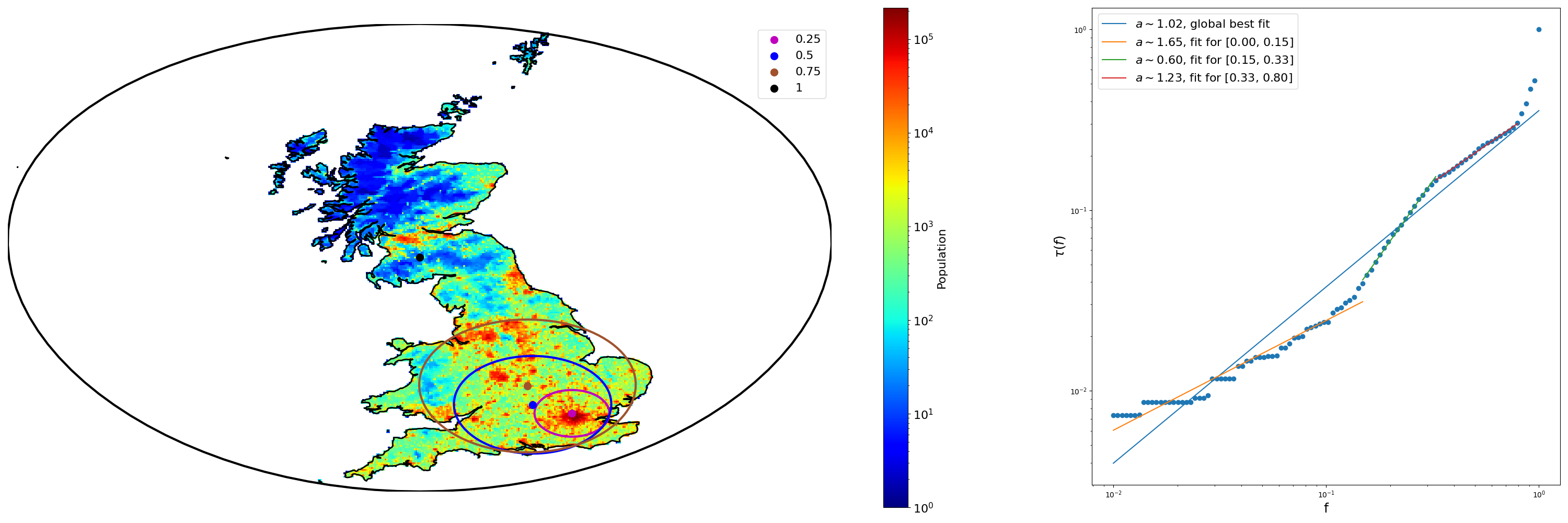}
    \caption{Left: VP 0.25-,0.5-,0.75- and 1-circles for Great Britain. Right: VP profile with best fit lines, note the log axes. Using SEDAC 2.5 minute resolution, $\sim 5 \text{km}$, data. }
    \label{fig:profuk}
\end{figure}
Figure \ref{fig:profuk} shows the VP profile for the United Kingdom as well as some of its VP f-circles. The global best fit corresponds to $a \sim 1$ though the graph of $\tau$ has some interesting inflection points indicating there is structure in the data not captured by the simple disc model. From $f=0$ to $f\sim0.15$ the VP circle is entirely contained in London and $\tau$ increases at a rate close to $a=2$. Once $f$ is large enough that we need to go outside of London, the slope changes abruptly to cover the gap between the densely populated areas of England in the south east and north west. Once the north west of England is reached, between $f \sim 0.33$ and $f \sim 0.8$ the radius increases more slowly until much bigger jumps are required to reach the remaining population in the north east of England and then Scotland.

\begin{figure}
    \centering
    \includegraphics[width=\textwidth]{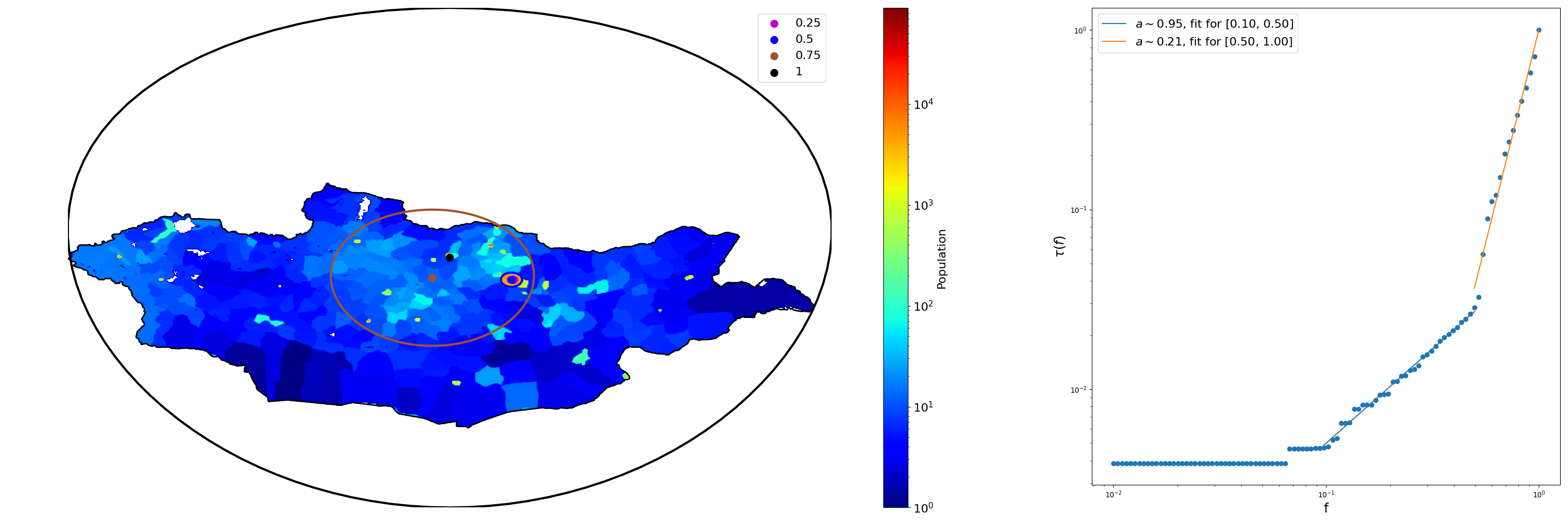}
    \caption{Left: VP 0.25-,0.5-,0.75- and 1-circles for Mongolia. Right: VP profile with best fit lines, note the log axes. Using SEDAC 2.5 minute resolution, $\sim 5 \text{km}$, data. }
    \label{fig:profmongolia}
\end{figure}
\begin{figure}
    \centering
    \includegraphics[width=\textwidth]{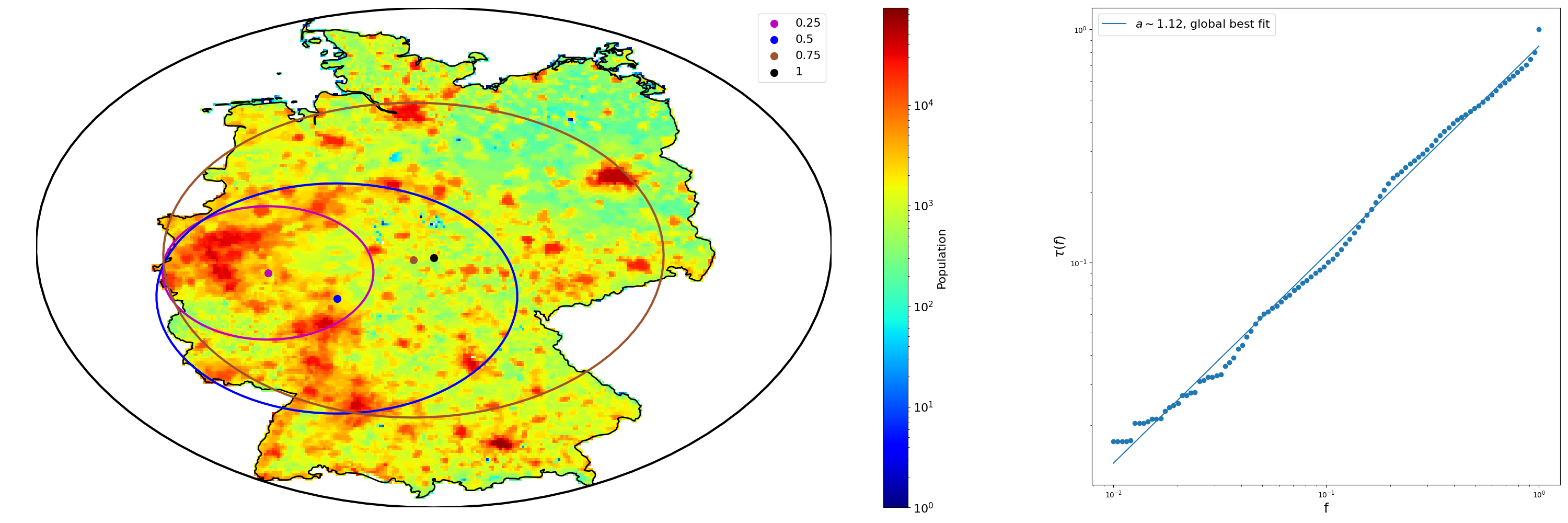}
    \caption{Left: VP 0.25-,0.5-,0.75- and 1-circles for Germany. Right: VP profile with best fit lines, note the log axes. Using SEDAC 2.5 minute resolution, $\sim 5 \text{km}$, data. }
    \label{fig:profnetherlands}
\end{figure}
Figure \ref{fig:profmongolia} shows Mongolia, a rather clear example, where around 50\% of the population live at high densities in the capital and the rest are extremely dispersed. This results in a sharp inflection point around $f \sim 0.5$. The opposite extreme is Germany, Figure \ref{fig:profnetherlands}, which has a comparatively uniform population distribution and is lacking obvious inflection points, so the disc model does a good job fitting this data. Even in cases (e.g. UK and Mongolia) where the global fit is bad, a model where the radial decay rate $a$ abruptly changes still fits the data well.

\begin{table}
\centering
\begin{tabular}{|c|c|c|c|}
\hline
Country & Year & $C_{50}$ \\ \hline \hline
Mongolia (max) & 2020 & 0.96 \\
Mongolia & 2000 & 0.73 \\ \hline
Japan & 2020 & 0.84 \\ \hline
United Kingdom & 2020 & 0.70 \\ \hline
Global & 2020 & 0.68 \\ \hline 
Average Nation & 2020 & 0.62 \\ 
Average Nation & 2000 & 0.61 \\ \hline
China & 2020 & 0.6 \\ \hline
Kazakhstan & 2020 & 0.48 \\ \hline
India & 2020 & 0.48 \\ \hline
Continental US & 2000 & 0.48 \\
Continental US & 2020 & 0.43 \\ \hline
Germany & 2020 & 0.35 \\ \hline
Sierra Leone (min) & 2020 & 0.27 \\ \hline
\end{tabular}
\caption{The centralisation $C_{50}$ metric for various countries mentioned previously as well as the max and min values for non-island nations (these have very large $R(1)$ values) found in the SEDAC 2020, 2.5 minute data.}
\label{tab:cent}
\end{table}
While the VP profile gives the most information and the slope parameter $a$ summarises some information about the population distribution, it is useful to have a single, model-independent, statistic to measure this tendency of areas to spread or concentrate their population. Define the statistic
\begin{equation}
C_{100f} = 1 - \frac{\tau(f)}{\sqrt{f}}
\end{equation} 
 as a measure of what I will call \textbf{centralisation}, I will use $C_{50}$ for simplicity. Large values (close to 1) correspond to highly centralised countries and smaller values to more dispersed ones. A uniform population on a disc gives $C_{50} = 0$ (the factor of $\sqrt{f}$ is to normalise this case). A population highly concentrated in the centre gives $C_{50} \simeq 1$. Table \ref{tab:cent} gives the centralisation scores for some countries of interest. The UK score is above average and the US score is below. The average nation's $C_{50}$ was relatively static between 2000 and 2020. As expected, Mongolia has a very high score and Germany a low one. Note that this is not simply a proxy for density, there are countries with low density e.g. Mongolia (2 people / km${}^2$) and Kazakhstan (7 people / km${}^2$) with both high and low $C_{50}$, 0.96 and 0.48 respectively. Likewise there are densely populated countries e.g. India (460 people / km${}^2$) and Japan (344 people / km${}^2$) with very different $C_{50}$, 0.48 and 0.84 respectively.

\section{Discussion}\label{sec:conc}

The VP circle (the standard version with $f=0.5$) is a kind of spatial median. Half of the people are inside the circle, the rest are outside. Like the one dimensional median, it is less affected by extreme values than the mean. Thus the VP centre tends to be closer to the main mass of the population, e.g. the north eastern US or south eastern UK. The VP radius provides a useful measure of population dispersion. Section \ref{sec:cpop} showed that this radius increases over time for the US likely due to disproportionate growth of population in the more sparsely populated west and south. Mongolia shows the opposite behaviour, population movement from rural to urban areas results in a rapidly decreasing radius. VP profiles, defined in Section \ref{sec:profiles}, provide a simple way to reduce two dimensional population maps to a one dimensional function while retaining much of the important information. A summary statistic, $C_{50}$, was also defined which measures the tendency of countries to spread their populations evenly or concentrate them in a central city or area. This seems to capture an interesting feature of population that is not measured by any other statistic of which I am aware.

Improvements and modifications to the definitions could of course be made. The algorithm described in Section \ref{sec:finding} could be sped up using some search heuristics, though this has not been necessary for this work, all of which was performed on a very modest laptop. Faster searches however might be necessary if the VP circle is generalised e.g. the VP ellipse (the ellipse of minimum area that encloses a fraction $f$ of the population) or the VP polygon (the polygon of minimum area what encloses a fraction $f$ of the population). Such generalisations could help for very non-circular areas e.g. Chile, but I leave this for future work. 

 As well as academic, there is significant popular interest in population centres. The Wikipedia page for center of population\footnote{  \url{https://en.wikipedia.org/wiki/Center_of_population} Accessed 28/07/23}, gives a dozen or more examples of population centres in different countries and there are specific pages for the center of the UK\footnote{ \url{https://en.wikipedia.org/wiki/Centre_points_of_the_United_Kingdom}Accessed 28/07/23} and at least five discussing various centres of the US\footnote{ \url{https://en.wikipedia.org/wiki/List_of_geographic_centers_of_the_United_States} Accessed 28/07/23}\footnote{ \url{https://en.wikipedia.org/wiki/Geographic_center_of_the_United_States} Accessed 28/07/23}\footnote{ \url{https://en.wikipedia.org/wiki/Mean_center_of_the_United_States_population} Accessed 28/07/23}\footnote{ \url{https://en.wikipedia.org/wiki/Median_center_of_United_States_population} Accessed 28/07/23}\footnote{ \url{https://en.wikipedia.org/wiki/Geographic_center_of_the_United_States} Accessed 28/07/23}. VP circles for other statistics could also be of academic or practical interest.  The VP circle of economic activity would be an area containing half the world's economic output. Following \cite{grether2010world} or \cite{quah2011global} this would be interesting with regards to tracking the relative growth of India and China versus the historical centres of Europe and the US. In the study of disease, the VP circle of infected people (the minimum area containing half of the infected population) could be a useful way to locate the epicentre and geographic extent of an outbreak and give a good idea of where to concentrate resources. This paper has aimed to show that the VP circle and statistics derived from it can be of use in the analysis of spatial data.
\bibliographystyle{plainnat}
\bibliography{main}

\begin{thebibliography}{25}
\providecommand{\natexlab}[1]{#1}
\providecommand{\url}[1]{\texttt{#1}}
\expandafter\ifx\csname urlstyle\endcsname\relax
  \providecommand{\doi}[1]{doi: #1}\else
  \providecommand{\doi}{doi: \begingroup \urlstyle{rm}\Url}\fi

\bibitem[Aboufadel and Austin(2006)]{aboufadel2006new}
Edward Aboufadel and David Austin.
\newblock A new method for computing the mean center of population of the
  {U}nited {S}tates.
\newblock \emph{The Professional Geographer}, 58\penalty0 (1):\penalty0 65--69,
  2006.

\bibitem[Bachi(1963)]{bachi1963standard}
Roberto Bachi.
\newblock Standard distance measures and related methods for spatial analysis.
\newblock \emph{Papers of the Regional Science Association}, 10\penalty0
  (1):\penalty0 83--132, 1963.

\bibitem[Barmore(1991)]{barmore1992we}
Frank~E Barmore.
\newblock Where are we? comments on the concept of the ``center of population".
\newblock \emph{The Wisconsin Geographer}, 7:\penalty0 40--50, 1991.

\bibitem[Beck and Sabach(2015)]{beck2015weiszfeld}
Amir Beck and Shoham Sabach.
\newblock Weiszfeld’s method: Old and new results.
\newblock \emph{Journal of Optimization Theory and Applications}, 164:\penalty0
  1--40, 2015.

\bibitem[Behnel et~al.(2011)Behnel, Bradshaw, Citro, Dalcin, Seljebotn, and
  Smith]{behnel2011cython}
Stefan Behnel, Robert Bradshaw, Craig Citro, Lisandro Dalcin, Dag~Sverre
  Seljebotn, and Kurt Smith.
\newblock Cython: The best of both worlds.
\newblock \emph{Computing in Science \& Engineering}, 13\penalty0 (2):\penalty0
  31--39, 2011.

\bibitem[Breau et~al.(2018)Breau, Toy, Brown, Macdonald, and
  Coomes]{breau2018footsteps}
S{\'e}bastien Breau, Brian Toy, Mark Brown, Ryan Macdonald, and Oliver~T
  Coomes.
\newblock In the footsteps of {M}ackintosh and {I}nnis: Tracking {C}anada’s
  economic centre of gravity since the great depression.
\newblock \emph{Canadian Public Policy}, 44\penalty0 (4):\penalty0 356--367,
  2018.

\bibitem[{Center for International Earth Science Information Network - CIESIN -
  Columbia University}(2018)]{sedac}
{Center for International Earth Science Information Network - CIESIN - Columbia
  University}.
\newblock Gridded population of the world, version 4.11 (gpwv4): Population
  count, revision 11., 2018.
\newblock URL \url{https://doi.org/10.7927/H4JW8BX5}.
\newblock [Online; accessed 24-July-2023].

\bibitem[Dorling and Atkins(1995)]{dorling1995population}
Daniel Dorling and David Atkins.
\newblock Population density, change and concentration in great britain 1971,
  1981 and 1991.
\newblock \emph{STUDIES ON MEDICAL AND POPULATION SUBJECTS NO.58}, 1995.

\bibitem[El~Deeb(2021)]{el2021spatial}
Omar El~Deeb.
\newblock Spatial autocorrelation and the dynamics of the mean center of
  covid-19 infections in lebanon.
\newblock \emph{Frontiers in Applied Mathematics and Statistics}, 6:\penalty0
  620064, 2021.

\bibitem[Fisher(2013)]{washingtonval}
Max Fisher.
\newblock 40 maps that explain the world.
\newblock
  \url{https://www.washingtonpost.com/news/worldviews/wp/2013/08/12/40-maps-that-explain-the-world/},
  2013.
\newblock [Online; accessed 24-July-2023].

\bibitem[{Geography Division U.S. Census Bureau}(2011)]{uscentres}
{Geography Division U.S. Census Bureau}.
\newblock Centers of population computation for the united states 1950 - 2010,
  2011.

\bibitem[Grether and Mathys(2010)]{grether2010world}
Jean-Marie Grether and Nicole~A Mathys.
\newblock Is the world's economic centre of gravity already in asia?
\newblock \emph{Area}, 42\penalty0 (1):\penalty0 47--50, 2010.

\bibitem[Hall et~al.(2019)Hall, Bustos, Ol{\'e}n, and
  Niedomysl]{hall2019population}
Ola Hall, Maria Francisca~Archila Bustos, Niklas~Boke Ol{\'e}n, and Thomas
  Niedomysl.
\newblock Population centroids of the world administrative units from nighttime
  lights 1992-2013.
\newblock \emph{Scientific Data}, 6\penalty0 (1):\penalty0 235, 2019.

\bibitem[{International Organization for Migration (IOM)}
  et~al.(2021){International Organization for Migration (IOM)}, of~Mongolia
  Population~Training, Centre, Economic, on~Innovation, and
  (UNU–MERIT)]{mongolia}
{International Organization for Migration (IOM)}, National~University
  of~Mongolia Population~Training, Research Centre, United Nations
  University~Maastricht Economic, Social Research~Institute on~Innovation, and
  Technology (UNU–MERIT).
\newblock \emph{Mongolia: Migration and Employment Study}.
\newblock IOM. Ulaanbaatar., 2021.

\bibitem[Murray(2018)]{murray2018complexities}
Alan~T Murray.
\newblock Complexities in spatial center derivation.
\newblock \emph{Transactions in GIS}, 22\penalty0 (6):\penalty0 1335--1350,
  2018.

\bibitem[Niedomysl et~al.(2017)Niedomysl, Hall, Archila~Bustos, and
  Ernstson]{niedomysl2017using}
Thomas Niedomysl, Ola Hall, Maria~Francisca Archila~Bustos, and Ulf Ernstson.
\newblock Using satellite data on nighttime lights intensity to estimate
  contemporary human migration distances.
\newblock \emph{Annals of the American Association of Geographers},
  107\penalty0 (3):\penalty0 591--605, 2017.

\bibitem[Plane and Rogerson(2015)]{plane2015tracking}
David~A Plane and Peter~A Rogerson.
\newblock On tracking and disaggregating center points of population.
\newblock \emph{Annals of the Association of American Geographers},
  105\penalty0 (5):\penalty0 968--986, 2015.

\bibitem[Quah(2011)]{quah2011global}
Danny Quah.
\newblock The global economy’s shifting centre of gravity.
\newblock \emph{Global Policy}, 2\penalty0 (1):\penalty0 3--9, 2011.

\bibitem[Quah(2015)]{quahtight}
Danny Quah.
\newblock The world’s tightest cluster of people.
\newblock
  \url{https://blogs.lse.ac.uk/maths/2015/12/16/the-worlds-tightest-cluster-of-people/},
  2015.
\newblock [Online; accessed 24-July-2023].

\bibitem[Quah(2016)]{quah2016ordering}
Danny Quah.
\newblock Ordering the world: Truth to power, 2016.

\bibitem[Rogerson(2015)]{rogerson2015new}
Peter~A Rogerson.
\newblock A new method for finding geographic centers, with application to {US}
  states.
\newblock \emph{The Professional Geographer}, 67\penalty0 (4):\penalty0
  686--694, 2015.

\bibitem[Rogerson(2021)]{rogerson2021historical}
Peter~A Rogerson.
\newblock Historical change in the large-scale population distribution of the
  {U}nited {S}tates.
\newblock \emph{Applied geography}, 136:\penalty0 102563, 2021.

\bibitem[Vardi and Zhang(2001)]{vardi2001modified}
Yehuda Vardi and Cun-Hui Zhang.
\newblock A modified {W}eiszfeld algorithm for the {F}ermat-{W}eber location
  problem.
\newblock \emph{Mathematical Programming}, 90:\penalty0 559--566, 2001.

\bibitem[Weiszfeld(1937)]{weiszfeld}
E.V. Weiszfeld.
\newblock Sur le point pour lequal la somme des distances de n points donnés
  est minimum.
\newblock \emph{Tohoku Math J.}, 43:\penalty0 335--386, 1937.

\bibitem[{World Bank}(2023)]{WBpoly}
{World Bank}.
\newblock World boundaries geojson - very high resolution.
\newblock
  https://datacatalog.worldbank.org/search/dataset/0038272/World-Bank-Official-Boundaries,
  2023.
\newblock Data retrieved 01/07/2023.

\end{thebibliography}

\end{document}